\documentclass{article}

\usepackage{cite}
\usepackage{mathtools}
\usepackage{amsmath}
\usepackage{amssymb}
\usepackage{hyperref}
\usepackage{cleveref}
\usepackage{bm}
\usepackage{gensymb}
\usepackage[margin=0.9in]{geometry}

\begin{document}

{\huge
\textbf\newline{Physical model simulator-trained neural network for computational 3D phase imaging of multiple-scattering samples}
}
\newline
\\
Alex Matlock\textsuperscript{1}*
 and Lei Tian\textsuperscript{1},
\\
{1. Department of Electrical and Computer Engineering, Boston University, Boston, MA 02215, USA}
\\
*amatlock@bu.edu

\begin{abstract}
Recovering 3D phase features of complex, multiple-scattering biological samples traditionally sacrifices computational efficiency and processing time for physical model accuracy and reconstruction quality.
This trade-off hinders the rapid analysis of living, dynamic biological samples that are often of greatest interest to biological research.
Here, we overcome this bottleneck by combining annular intensity diffraction tomography (aIDT) with an approximant-guided deep learning framework.
Using a novel physics model simulator-based learning strategy trained entirely on natural image datasets, we show our network can robustly reconstruct complex 3D biological samples of arbitrary size and structure.
This approach highlights that large-scale multiple-scattering models can be leveraged in place of acquiring experimental datasets for achieving highly generalizable deep learning models.
We devise a new model-based data normalization pre-processing procedure for homogenizing the sample contrast and achieving uniform prediction quality regardless of scattering strength.
To achieve highly efficient training and prediction, we implement a  lightweight 2D network structure that utilizes a multi-channel input for encoding the axial information. 
We demonstrate this framework's capabilities on experimental measurements of epithelial buccal cells and {\it Caenorhabditis elegans} worms.
We highlight the robustness of this approach by evaluating dynamic samples on a living worm video, and we emphasize our approach's generalizability by recovering  algae samples evaluated with different experimental setups.
To assess the prediction quality, we develop a novel quantitative evaluation metric and show that our predictions are consistent with our experimental measurements and multiple-scattering physics.
\end{abstract}

\section{Introduction}
\label{section:intro}
Complex biological sample 3D recovery remains an outstanding difficulty in quantitative phase imaging (QPI) due to the inherent trade-off in traditional techniques between physical model accuracy and computational efficiency.
The most efficient 3D QPI techniques utilize single-scattering models, such as the first Born or Rytov approximations, providing computationally efficient closed-form solutions for volumetric recovery~\cite{jin.etal2017, park2018quantitative}.
While both interferometry-based~\cite{jin.etal2017, park2018quantitative} and intensity-based~\cite{rodrigo2017, jenkins2015,ling2018high, li2019high,matlock_2019} 3D QPI methods using these approximations are successful in applications including immuno-oncology~\cite{nandakumar2012isotropic}, cytopathology~\cite{park2008}, and stem cell research~\cite{sandoz2018label}, the single-scattering model underestimates the refractive index (RI) of multiple-scattering samples~\cite{li2019high, lim2019high, chen2020multi} preventing the quantitative analysis of complex biology, such as tissue biopsies and organoids. 
Recent efforts have shown improved RI estimates can be achieved using multiple-scattering model based iterative reconstruction algorithms~\cite{tian20153d,kamilov2015learning, chowdhury2019high,lim2019high, tahir2019holographic, chen2020multi}.
However, this improved accuracy requires greater computation times that limit the evaluation of dynamic samples and large-scale objects in time-series studies.

\begin{figure}[t!]
\centering
\includegraphics[width =1\textwidth]{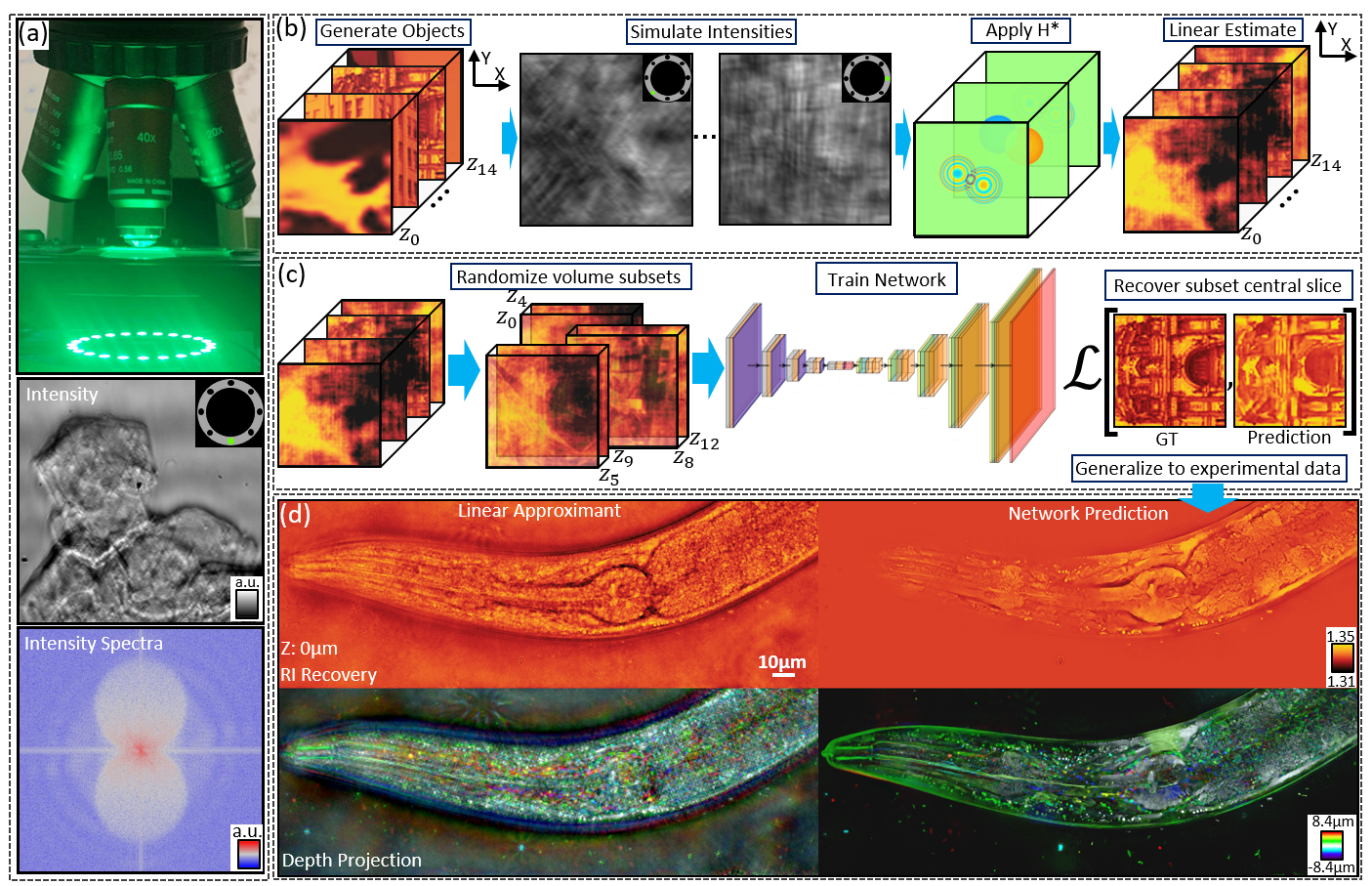}
\caption{Overview of the proposed physical model simulator-trained neural network for computational 3D phase imaging.
	(a) annular intensity diffraction tomography (aIDT) imaging setup (top) with example intensity image (middle) and intensity spectra (bottom) under single-LED oblique illumination. 
	(b) Simulation process for generating network training data. Natural images are randomly sorted into volumes with randomized RI, the intensity images matching the aIDT setup are simulated using the SSNP multiple-scattering forward model, and linear approximants are generated using the aIDT model for the network inputs.
	(c) Training process for the proposed neural network. The simulated object volumes are randomly segmented into five slice subsets on each mini-batch during training and fed into the network and the central volume slice is recovered.
	(d) Example application of the network on an experimentally measured {\it C. elegans} worm compared with the aIDT linear object estimate with in-focus RI slice reconstructions (top) and color-coded depth projections (bottom). Results demonstrate network generalization, enhanced depth sectioning, and improved feature recovery using the proposed network.
}
\label{F1}
\end{figure}

Here, we overcome this bottleneck between efficiency and accuracy by melding closed-form single-scattering solutions with a fast and generalizable deep learning model.
We illustrate this synergistic approach on our recently developed annual intensity diffraction tomography (aIDT) system, a computational 3D phase imaging modality using oblique illumination from a ring LED illuminator to encode 3D phase into intensity measurements (Fig.~\ref{F1}(a))~\cite{li2019high}. 
Our prior work demonstrated efficient 3D phase recovery with 10.6 Hz volume rates based on a linear single-scattering model (Fig.~\ref{F1}(a))~\cite{li2019high}.
To maintain fast and quantitative volumetric 3D phase imaging, we combine aIDT with the proposed deep learning model for live sample imaging with minimal artifacts on multiple-scattering specimens.

Deep learning (DL) has revolutionized the fields of computational microscopy with its ability to efficiently solve complex nonlinear inverse problems~\cite{barbastathis2019use}.
Existing DL models utilize different learning strategies from full ``end-to-end'' models for direct inversion~\cite{sinha2017lensless, xue2019reliable} to ``guided'' learning  with embedded physical models or priors~\cite{kellman2019physics, lim2020three, wu2020simba, zhou2020diffraction, kang2020limited, rivenson2018phase}.
The physics-guided approach is appealing as it encourages predicting physics-constrained features and reduces the ``black box'' nature of DL models.
While many DL QPI developments have been in 2D~\cite{sinha2017lensless, xue2019reliable}, recent works have expanded to 3D QPI using the physical approximant-guided learning approach~\cite{lim2020three, kang2020limited, goy201821378}.
These approaches successfully improved RI predictions on red blood cells~\cite{lim2020three} and high-contrast manufactured samples~\cite{kang2020limited, goy201821378}.
However, the generalizability of existing networks are limited due to the similarities between the training and testing data.
For biological applications where objects of interest may vary significantly between specimen types, the potential for overfitting in existing approaches significantly limits their broad application.
Furthermore, all existing networks for 3D QPI utilize 3D network structures and contain a large number of trainable parameters that further complicate the training data size and computational requirements for tomographic recovery. 

Here, we overcome these existing limitations by leveraging multiple-scattering model simulation, efficient network architectures, and single-scattering approximant to achieve efficient multiple-scattering object recovery on dynamic biological samples. 
First, we develop a physical model simulator-based training approach bypassing the need for experimentally acquiring diverse training datasets with accessible ground truth information.
To facilitate multiple-scattering feature recovery, we generate simulated object volumes using a fast and accurate Split-Step Non-Paraxial (SSNP)~\cite{lim2019high} multiple-scattering model (Fig.~\ref{F1}(b)).
We enforce model generalizability with these object volumes by generating them from diverse natural images available from multiple open-source databases.
Using aIDT’s single-scattering based 3D estimates of these objects as the network’s input, we train the network to enhance the 3D RI recovery and correct model-based artifacts using an approximant-guided learning strategy.
Second, our network features a lightweight 2D ``U-Net'' structure to perform 3D reconstruction~\cite{ronneberger2015u}. 
We achieve efficient learning with this network by feeding five consecutive axial slices selected randomly from larger object volumes as a multi-channel input and predict only the central object slice (Fig.~\ref{F1}(c)). 
We show this approach efficiently encodes the depth and diffraction information and enables effective suppression of missing-cone and multiple-scattering artifacts in a highly scalable and computational efficient manner as compared to alternative 3D networks or other complex architectures~\cite{lim2020three, kang2020limited}.
Third, to provide uniform prediction quality regardless of scattering strength, we devise a model-based data normalization procedure for homogenizing sample contrast prior to the model prediction.
This novel data preprocessing procedure dramatically improves the model's generalizability in terms of RI contrast. 
In combination, we show that our network can be generalized to recover complex 3D biological samples of arbitrary size and structure.

We experimentally demonstrate our network's superior generalization capacity by predicting live epithelial buccal cells, {\it Caenorhabditis elegans} worm samples (Fig.~\ref{F1}(d)), and fixed algae samples acquired using different experimental setups (Supplemental).
We further highlight the robustness of our network by making time-series predictions on a living, dynamic worm.
To quantitatively assess the reliability of our network's predictions, we adapt an image-space based evaluation procedure by feeding the network predicted RI into the multiple-scattering model and comparing the calculated intensity and experimental measurements. 
Even for ``unseen'' illumination angles that are unused during model training or prediction, the calculated intensities from our predictions match well with experimental data.
We show a 2-3$\times$ error reduction using our network over the aIDT linear model's estimates.
Our result highlights that leveraging large-scale multiple-scattering modeling can obviate major overhead in physical data acquisition and train a reliable, highly generalizable deep learning model for imaging complex 3D biology.

\section{Network architecture for optimal recovery}
\label{section:NetworkArchitecture}
To optimally combine aIDT's reconstruction pipeline with a learning model, the network architecture must satisfy four key properties: 1) preserve the modality's speed, 2) provide arbitrary volume size recovery, 3) remove single-scattering approximation and missing-cone induced artifacts, and 4) equivalently recover weak and highly scattering samples.
The first two properties preserve the main benefits from the aIDT platform and linear model, while the latter two properties require improved performance over aIDT's model-based implementation.
These factors require the network to learn efficient object predictions robust to scattering strengths without sacrificing aIDT's fast acquisition speeds.

\begin{figure}[t!]
\centering
\includegraphics[width =1 \textwidth]{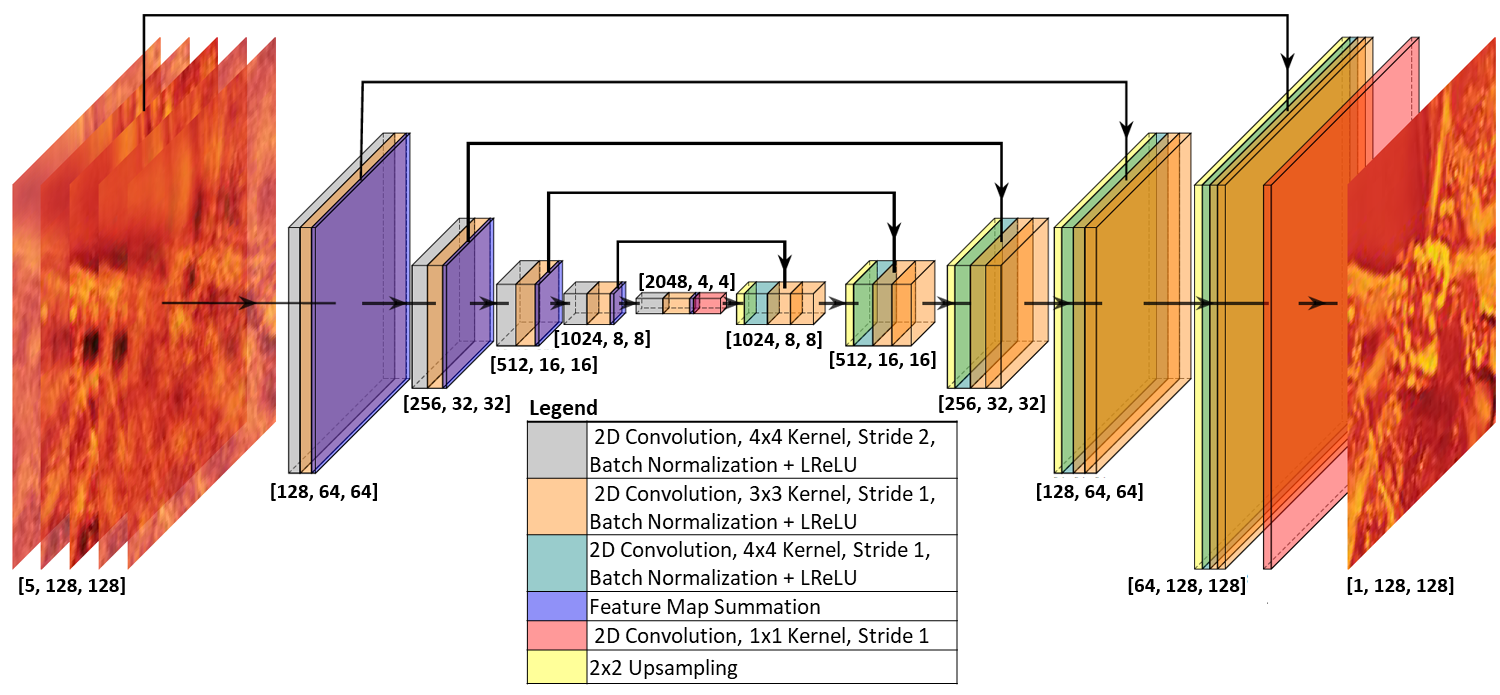}
\caption{Proposed modified 2D U-Net architecture. Five consecutive axial slices from the aIDT linear approximant are fed into the network and the central slice is predicted on the output.
Architecture details are provided in section~\ref{section:MM}.
Network visualization generated using~\cite{haris_iqbal_2018_2526396}.
}
\label{F2}
\end{figure}

Satisfying the defined properties required specific choices in the data generation, training process, and network architecture as shown in Figs. \ref{F1} and~\ref{F2}, respectively.
In generating a robust network for evaluating arbitrary object samples under supervised learning conditions, a significant constraint exists in acquiring sufficiently diverse experimental measurements to properly train a network without overfitting to specific object types.
This limitation often restricts networks to recovering specific biological sample types~\cite{lim2020three, kang2020limited}.
Here, we completely bypass this issue by {\it simulating} object volumes for training the network from diverse, readily available open-source natural image datasets~\cite{xiao2010sun, huang2008labeled, griffin2007caltech, cheng2017remote, oh2016deep, krizhevsky2009learning, khosla2011novel}. 
Randomly selecting natural images from these datasets, we stack these images into 3D volumes and assign them random RI values to create weak and strongly scattering media.
We then leverage a highly efficient and accurate Split-Step Non-Paraxial (SSNP) forward model~\cite{lim2019high} to generate intensity images with the same physical parameters as the experimental aIDT setup (Fig.~\ref{F1}(b))~\cite{li2019high}.
This simulator provides a rigorous physical model of the multiple-scattering through an object using a multi-slice beam propagation-based approach accounting for both the scattered field and its first-order axial derivative at each object slice~\cite{lim2019high}.
Using SSNP allows easy, rapid generation of a large diverse dataset for training the network while the use of natural images helps prevent overfitting from their high entropy~\cite{deng2020interplay}.
With these images, we can obtain linear model approximants of the object volumes using the aIDT model, which we use as the network inputs for training.
Details on the simulation process for this dataset can be found in Section \ref{section:MM}.

With this simulated training dataset, we next develop a training process to help the network recover volumes of arbitrary RI contrast and size.
A significant challenge in recovering both weak and strongly scattering objects with a single learning model is the heterogeneity of the data distribution.
When training directly on the simulated dataset, we observe that the larger approximant error from strongly scattering objects results in network overfitting to correct high scattering strength features while over-smoothing weak scattering structures.
We overcome this issue with a linear model-based data normalization scheme, detailed in Section \ref{section:MM}, to homogenize the dataset and enable high-quality object predictions regardless of scattering strength.

To enable arbitrary volume prediction, we introduce a randomized sampling procedure to the training process.
Due to the sparse illumination and undersampling of the object's Ewald sphere in aIDT's limited angle tomography design, the object's linear approximant exhibits anisotropic, axially varying, and object-dependent missing cone artifacts throughout the volume.
As a result, these artifacts exhibit unique behavior for each object and axial slice that necessitate a network to learn the entire linear model's 3D system transfer function.
To facilitate this process, we present random consecutive five slice subsets of an object as the input within each training mini-batch and predict the central slice from each subvolume (Fig.~\ref{F1}(c)) using a modified 2D U-Net (Fig.~\ref{F2})~\cite{ronneberger2015u}.
This procedure presents 3D information as the multi-channel input (i.e. feature maps), from which the network can extract 3D information, such as the system's 3D system transfer function and object's structure, with an efficient and easily trainable 2D network.
By randomizing the subvolume input during training, the network is forced to learn object recovery only from relative inter-slice relations allowing for arbitrary volume predictions.
Compared to a full 3D network, this approach provides nearly equivalent results with a 2.5$\times$ improvement in training time from its use of efficient 2D convolutions.
Details on the training procedure for this network and its architecture can be found in Section ~\ref{section:MM}.
Once trained, this network can be sequentially applied through the entire object volume to perform 3D predictions and has been tested on both unseen simulated objects (Supplemental) and experimental data such as the {\it Caenorhabditis elegans} ({\it C. elegans}) worm in Fig.~\ref{F1}(d).

\section{Results}

\subsection{Weakly scattering object recovery}
\label{subsection:WeakScatt}
\begin{figure}[t!]
\centering
\includegraphics[width =1 \textwidth]{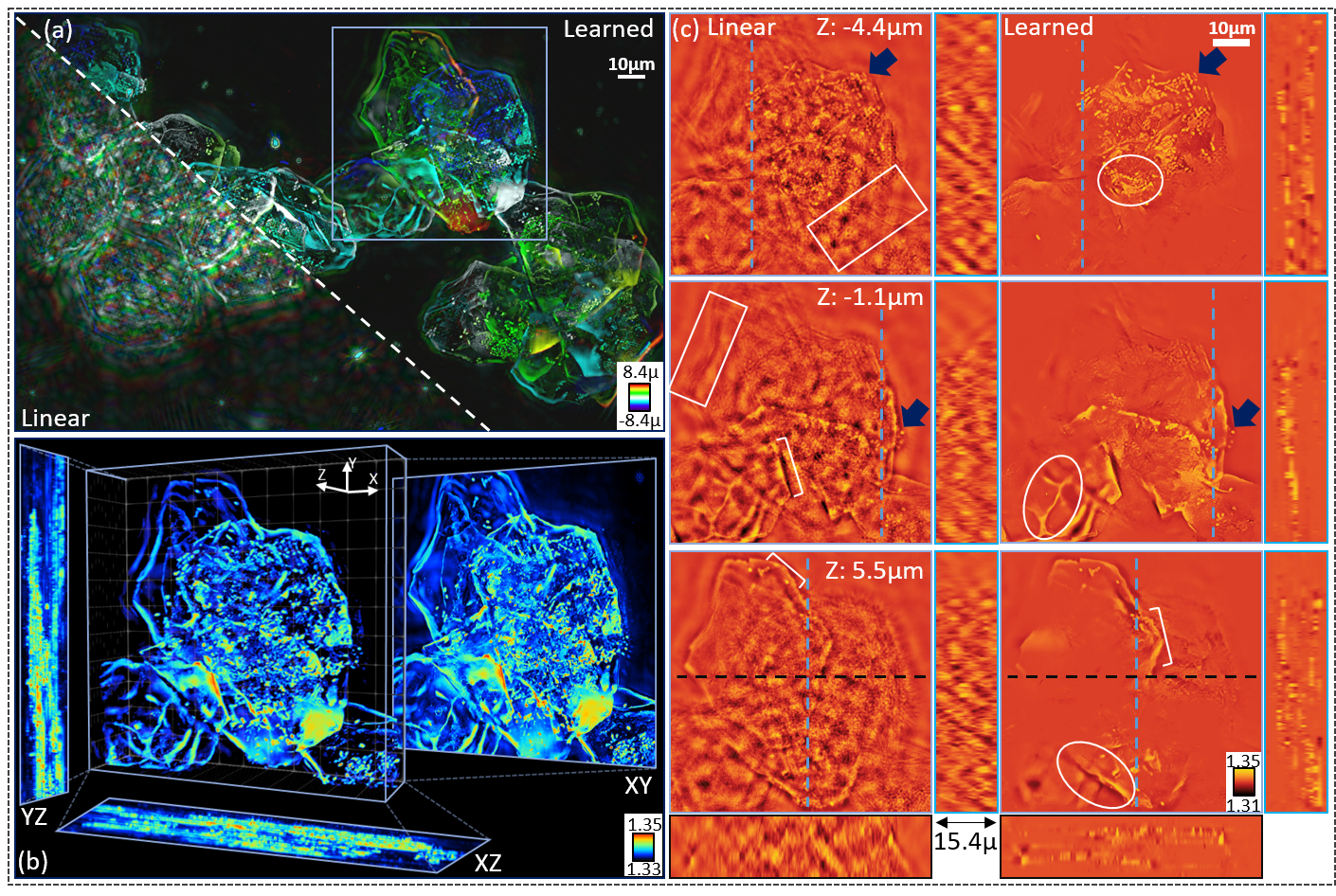}
\caption{Prediction results of weakly scattering epithelial buccal cells. 
	(a) Color-coded depth projections using the aIDT linear model (lower left) and learned result (upper right).
	(b) 3D rendering of the learned reconstruction outset from the purple region in (a). 
	(c) RI slice reconstructions, XZ, and YZ cross-sections from the linear (left) and learned (right) volume predictions. White squares show poor depth sectioning in the linear model that is corrected with the predicted results, while blue arrows highlight native bacteria features and white circles show enhanced cell edge detection in the learned result.
}
\label{F3}
\end{figure}

We first applied our network to weakly scattering epithelial buccal cell clusters in aqueous media ($n_0 = 1.33$) (Fig.~\ref{F3}).
Figure~\ref{F3}(a) compares the linear (lower left) and learned (upper right) depth-coded projections of the cell cluster volumes with outset comparisons of the RI at different slices in Fig.~\ref{F3}(c).
Fig.~\ref{F3}(b) shows a volume rendering of a cell cluster segment from our network with maximum intensity projections (MIP) in XY, YZ, and XZ. 

Compared with the linear case, the learned object prediction shows significant noise suppression and object feature enhancement.
Shown in the projection of Fig.~\ref{F3}(a) and the cross-sections of Fig.~\ref{F3}(c), the linear model generates strong missing cone artifacts corrupting the  reconstructed features.
While lateral cross-section images show cell edges (Fig.~\ref{F3}(c), white brackets) and native bacteria (Fig.~\ref{F3}(c), blue arrows) are visible with the linear estimate, the missing cone artifacts reduce feature visibility and confound the true morphology of cellular structures (Fig.~\ref{F3}(c), white boxes).
In contrast, the learned result maintains or improves recovery of these biological features (Fig.~\ref{F3}(c), white circles) and removes the model-induced missing cone artifacts as shown in the XZ and YZ cross-sections.
These improvements provide clearly distinguishable 3D volumes of the cellular structure (Fig.~\ref{F3}(b)) allowing easier evaluation of the sample's morphology.
Furthermore, the similar recovered RI values of the cell's edges and bacteria between the linear estimate and network prediction suggests the network recovers the correct quantitative values in weakly scattering media.
We further evaluate this quantitative recovery in section~\ref{subsection:Reliability}.
These results provide encouraging evidence that our network generalizes well to recovering weakly scattering, experimentally measured biological samples of arbitrary size, structure, and contrast.

We also evaluated the generalization capabilities of our network in predicting object volumes within different imaging mediums and a different optical setup (Supplemental).
These volumes consisted of a diatom sample embedded in glycerin ($n_0=1.45$) and a {\it spirogyra} sample measured using a low magnification (NA = 0.25, $M$ = 10) aIDT configuration.
The first sample changes the assumed imaging medium from the water medium assumed in the training dataset, while the second sample changes the physical imaging setup. 
These factors alter the missing cone artifact strength and structure throughout the volume and present unseen physical modifications that test whether the network generalizes to recovering enhanced object volumes under different imaging conditions.
As we discuss in the supplemental material, the network also recovers these structures well regardless of these varied object conditions and show our network is highly generalized.

\subsection{Multiple-scattering object recovery}
\label{subsection:StrongScatt}
\begin{figure}[t!]
\centering
\includegraphics[width =1 \textwidth]{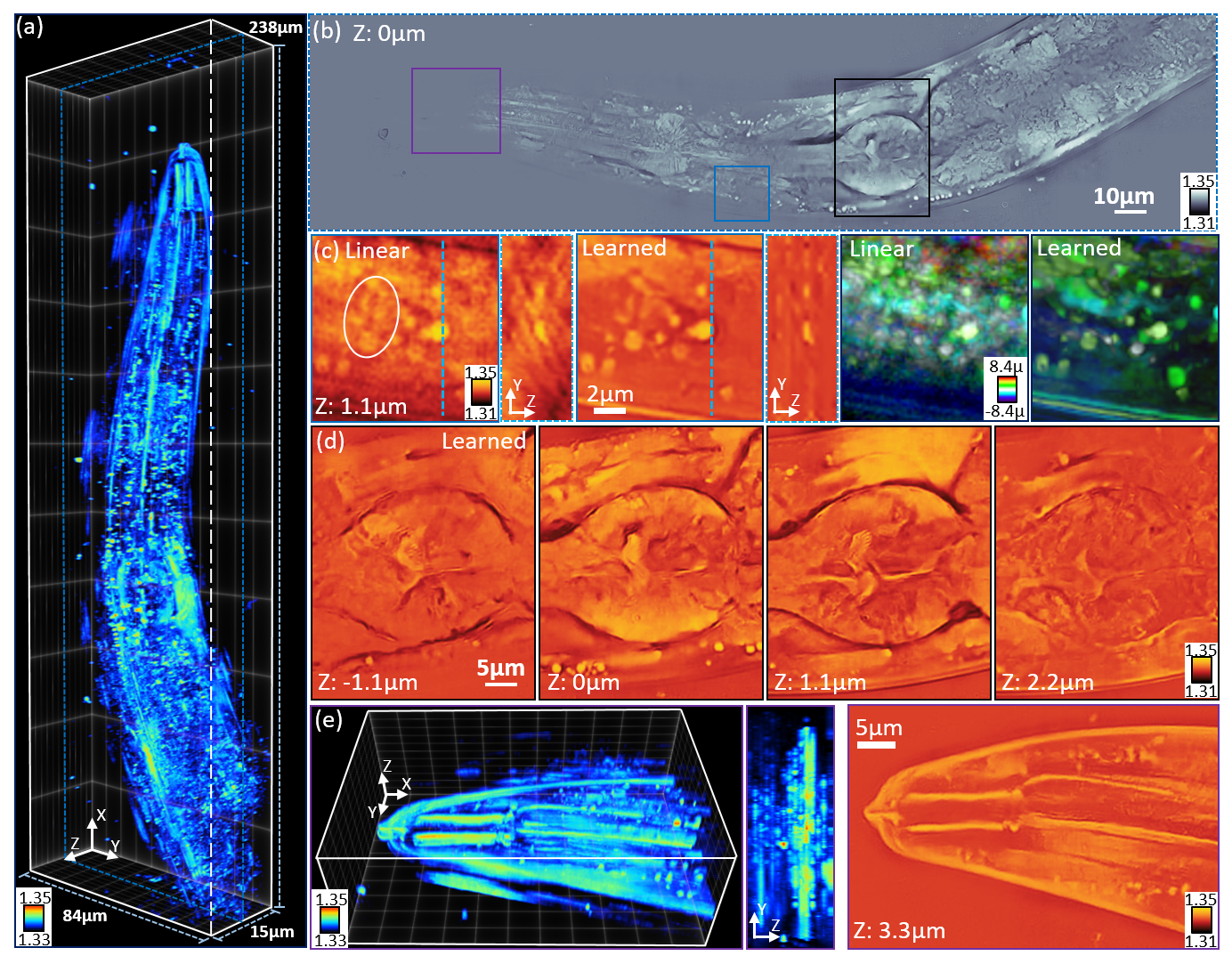}
\caption{Prediction results of multiple-scattering {\it C. elegans} worm sample.
	(a) Volume rendering of the prediction showing the full volume network recovery.
	(b) central slice reconstruction with outsets of lipid droplets (light blue), pharyngeal bulb (black) and buccal cavity (purple).
	(c) Outset comparing the linear aIDT reconstruction and learned prediction of lipid droplets in the sample with RI slice, YZ cross-sections, and color-coded projections.
	(d) Consecutive axial slices of the terminal pharyngeal bulb with clear recovery of the worm's grinder organ. High-resolution features are recovered with our network at 1.1$\mu$m,
	(e) Rendering, maximum intensity projection along YZ plane, and RI slice of the worm's buccal cavity. Results show features at defocused planes are well recovered with our network.
}
\label{F4}
\end{figure}

To evaluate the network's capabilities on stronger scattering media, we applied our learned model to a {\it C. elegans} worm sample as shown in Figure~\ref{F4}. 
Here, the figure shows a 3D rendered worm segment (Fig.~\ref{F4}(a)) with the central recovered RI slice (Fig.~\ref{F4}(b)) and outsets of tissue structures including lipid droplets (Fig.~\ref{F4}(c)), the terminal pharyngeal bulb with grinder (Fig.~\ref{F4}(d)), and the buccal cavity (Fig.~\ref{F4}(e)).
Immediately apparent in the learned prediction of the worm is the enhanced clarity and RI contrast of the various worm tissue structures.
The network's removal of missing cone artifacts and improved RI prediction show clean recovery of worm features across the entire segment in Fig.~\ref{F4}(b) and additional lipid droplets being recovered in Fig.~\ref{F4}(c).
Our learned approach further shows fine, continuous features are recoverable through the volume such as the grinder from the worm's digestive tract (Fig.~\ref{F4}(d)) and the pharyngeal epithelium at defocused planes in Fig.~\ref{F4}(e). 
While these features are also recovered using aIDT's linear model~\cite{li2019high}, the network's artifact removal and enhanced feature recovery significantly improves the depth sectioning of the reconstruction.
This is particularly evident with the buccal cavity centralized at 5$\mu$m whose missing cone artifacts have been nearly completely removed from the central slice (Fig.~\ref{F4}(a) purple square, ~\ref{F4}(e)).
These results highlight the network's capabilities to generalize well on multiple-scattering multi-cellular organisms.

\subsection{Dynamic sample volumetric recovery}
\label{subsection:Dynamic}
\begin{figure}[t!]
\centering
\includegraphics[width =1 \textwidth]{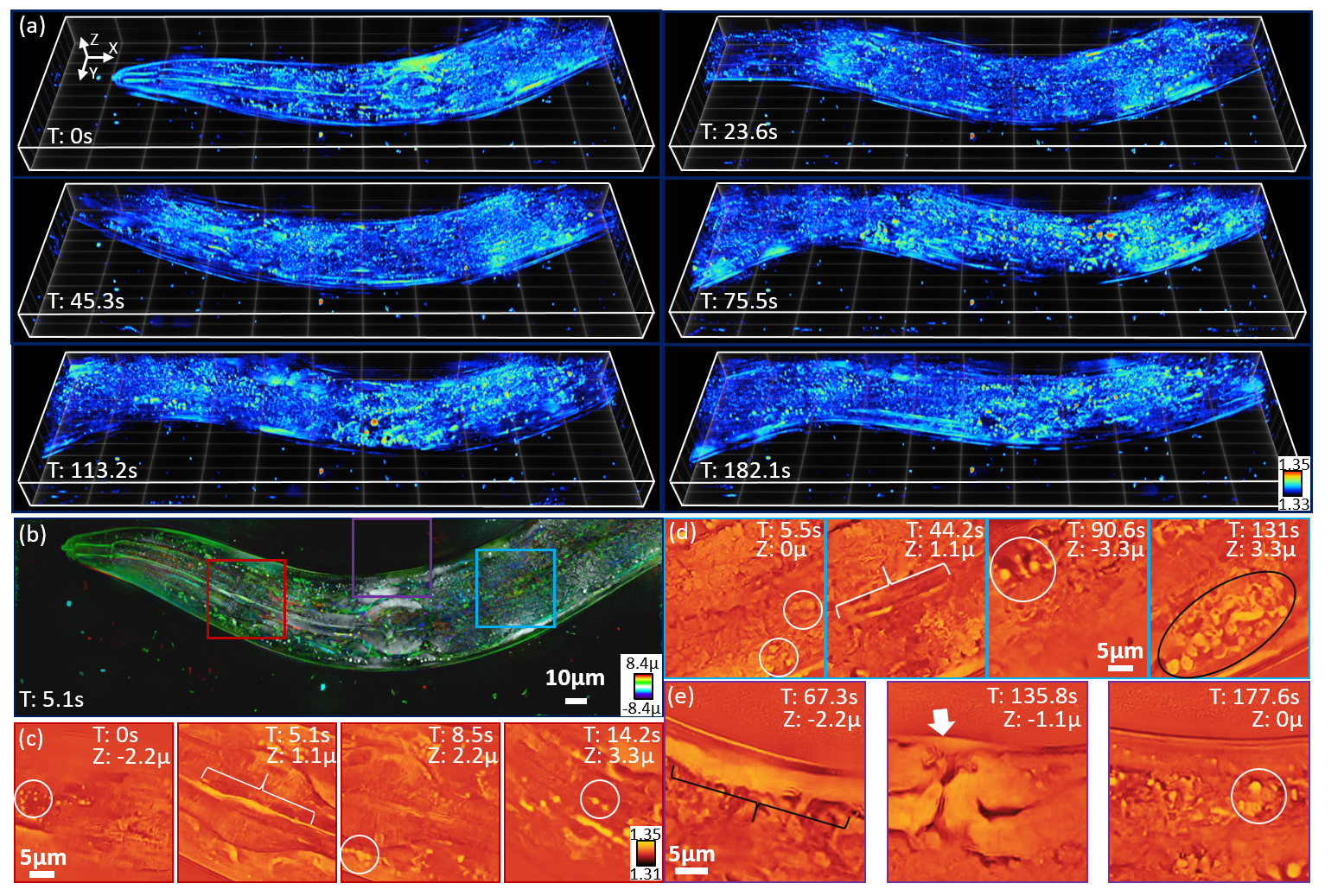}
\caption{Predictions of a dynamic {\it C. elegans} worm.
	(a) Learned object prediction 3D renderings across different time points of an aIDT longitudinal measurement.
	(b) Color-coded depth projection through the learned reconstruction at 5.1 seconds. Network prediction shows minimal missing cone artifacts and clear feature recovery,
	(c)-(e) outsets of reconstruction FOV highlighting recovered {\it C. elegans} organs and tissues during video reconstruction. White circles highlight lipid droplets and high-resolution circular structures, white brackets illustrate recovered intestinal tract, black brackets show worm muscle wall, black circles indicate complex tissue features being recovered, and the white arrow indicates the worm's vulva and reproductive organs.
}
\label{F5}
\end{figure}

One of the key objectives for combining aIDT with learning architectures is to maintain fast reconstruction of complex samples for imaging living dynamic biological samples.
To demonstrate this capability, we applied our trained network to {\it C. elegans} time-series measurements from~\cite{li2019high}.
Results are shown for specific time points in Figure~\ref{F5} and the video can be found in the supplemental material.

Figure~\ref{F5} highlights the wealth of information recovered by the network from the complex, dynamic biological samples.
From Fig.~\ref{F5}(a), the network predictions' removal of missing cone artifacts provides clear visualizations of the worm's movement through the entire 3-minute measurement period. 
This artifact removal also enhances the depth sectioning capabilities as seen previously, which is particularly evident in the well-separated features in the color-coded depth projection of Fig.~\ref{F5}(b).
During this time period, the learned model provides recovery of the digestive tract (white brackets) and lipid droplets (white circles) in Fig.~\ref{F5}(c) and Fig.~\ref{F5}(d) with complex internal organ features clearly recovered in Fig.~\ref{F5}(d) (black oval).
Figure~\ref{F5}(e) shows new feature recovery previously outside of the FOV including muscle walls and the worm's vulva (white arrow).

The network's enhanced recovery of such features in temporal data highlights its utility for arbitrary dynamic sample imaging.
Despite training on simulated natural images, the network's generalization recovers a complex biological sample's features consistently in time with minimal feature degradation.
This result opens the possibility for this network's application to evaluating temporal dynamics of biological samples with significantly enhanced feature recovery over conventional model-based aIDT.
This possibility highlights the significant potential of our learned approach to aIDT.

\subsection{Prediction reliability analysis}
\label{subsection:Reliability}

\begin{figure}[t!]
\centering
\includegraphics[width =1 \textwidth]{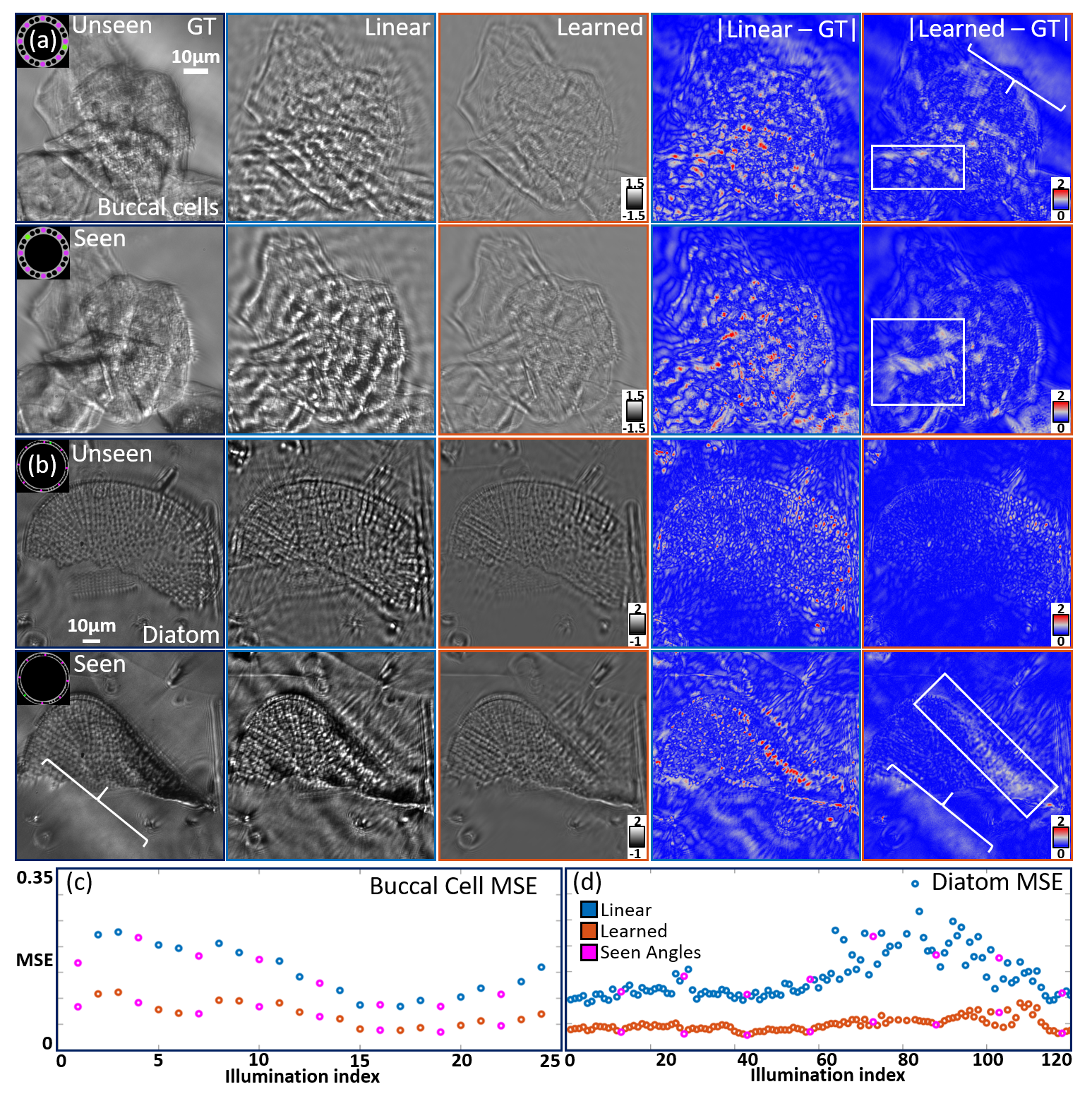}
\caption{Quantitative reliability analysis of network's predictions. 
We compare intensity images computed using the multiple-scattering model from the linear and learned object and compare them with the experimentally measured intensity images for (a) epithelial buccal cells and (b) a diatom sample. 
The learned object predictions show closer intensity image contrast and lower errors than intensity images computed from the IDT linear model-based reconstructions indicating closer object predictions to the ground truth.
	The MSE between the computed and experimental measured images across both seen and unseen angles for (c) epithelial buccal cells and (d) a diatom sample. 
	The results show consistently lower error using the learned model regardless whether the illuminations are used in the model training.
}
\label{F6}
\end{figure}

While the network predictions on simulated objects show close agreement with the ground truth (Supplemental), the network's reliability when predicting experimental data remains an outstanding question.
Despite replicating our aIDT setup's experimental parameters in simulation, variations in experimental measurements such as noise, illumination angle, source homogeneity, imaging wavelength, and aberrations could introduce artifacts to aIDT's object approximant.
These variations could generate unreliable object predictions that cannot be evaluated due to the lack of ground-truth information in experimental measurements.
Understanding the network reliability is crucial for applying this pipeline in biology where artificial features could cause mis-classification of features and/or disease mis-diagnoses.
To investigate this issue, we developed an image space analysis metric to evaluate the reliability of our network's predictions on experimental data.

Our image space metric expands upon the method in~\cite{lim2020three}.
This approach implements a physical forward model simulator to generate intensity images from the linear and learned object volumes that are then compared with the experimental measurements.
With a sufficiently rigorous physical simulator, deviations between the simulated and experimental images can be related to errors in the predicted RI of the recovered volumes.
This comparison utilizes the ground-truth object information encoded in the experimental measurements to evaluate the predicted object estimates quantitatively.
Here, we expand upon this analysis by comparing intensity images from illumination angles used in our model-based approximant (seen) and illumination angles unused in the aIDT model and network prediction (unseen).
If the network ``overfits'' to the features recovered using seen illuminations, the simulated images from {\it unseen} illuminations would exhibit increased error from network hallucinated object features.

We first evaluate epithelial buccal cells using 24 illuminations in Fig.~\ref{F6}(a) and (c) using the setup described in Section~\ref{section:MM}.
To further evaluate the effect of experimental setup variations with this metric, we compute the metrics on diatom algae samples measured with a different LED array setup (Adafruit, 607) with 120 illuminations, and {\it spirogyra} samples using a low NA objective and a 36 LED ring array (details about the setup is provided in the Supplemental) in Fig.~\ref{F6}(b) and (d).
We utilize the SSNP model for generating the intensity images using both the linear (Fig.~\ref{F6}, blue) and learned (Fig.~\ref{F6}, orange) estimates and compared them with experimental measurements using pixel-wise absolute error maps (Fig.~\ref{F6}(a)) and the mean squared error (MSE) of each intensity image as a function of the illumination index (Fig.~\ref{F6}(b)). 
Seen illuminations are noted in magenta with green illuminations highlighting the specific LED used for the intensity images shown in Fig.~\ref{F6}(a).

The network's object predictions show strong agreement with the experimental measurements regardless of the illumination angle.
In both the cells and diatom sample, the learned network intensity images show closer contrast and lower error to the experimental data than the linear model's results.
This result is consistent regardless of whether the illumination angle was used for the reconstruction (Fig.~\ref{F6}(a), Seen vs. Unseen).
The main differences between the network and experimental images appears due to low spatial frequency loss (Fig.~\ref{F6}(a), white boxes) creating ``flatter'' images and source inhomogeneities in the experimental measurements
(Fig.~\ref{F6}(a), white brackets).
These error contributions are attributed to the linear model input to the network lacking low spatial frequency information and the LED sample illumination not ideally matching plane wave illumination, respectively.
These issues, however, are not tied to the network's behaviors and show the network is not introducing hallucinations or artifacts that would skew predictions of experimental objects.

Evaluating the image-wise MSE in Fig.~\ref{F6}(b) further confirms the network predicts the underlying object volume.
Plotted as a function of the illumination index, the MSE for the seen and unseen illumination angles show no substantial difference in error for both sample types.
Across all illuminations, the images from the learned object volume prediction show consistently lower error than the linear model, which is partially attributed to the removal of missing cone artifacts in the predicted volume.
The dominant error variations of Fig.~\ref{F6}(b) instead result from illumination angle misalignments remaining after implementing the calibration procedures of~\cite{eckert_2018, li2019high}.
These misalignments are most present in illuminations 70-90 for the diatom sample where the illumination angles were close to the objective's maximum 0.65 NA cut-off and were difficult to calibrate.
While these angle-based error fluctuations are most present in the aIDT model-based results, the network shows a stable MSE regardless of the illumination angle.
This stability is attributed to the removal of missing cone artifacts, as these features would generate significant error due to small fluctuations in the illumination angle.
These results highlight that our learned IDT framework provides reliable object estimates without network-induced hallucinations.

\section{Discussion and Conclusion}

Our results highlight the significant potential of deep learning in computational 3D phase imaging. 
With only simulated objects, we showed a lightweight 2D network can be trained following approximant-guided learning methods to recover the 3D phase of biological samples of arbitrary size and scattering strength.
The network corrects not only missing cone artifacts but also improves the prediction accuracy of the object's RI.
We showed better RI and object feature recovery on unseen simulated data and illustrated improved object predictions on experimental data acquired using a range of experimental setups. 
Finally, we showed this network can be readily applied to recover high-quality volumetric reconstructions of dynamic biological samples using a {\it C. elegans} worm video (Supplemental). 

A main limitation of this approximant-guided learning approach is the network's reliance on the initial model-based object estimate for feature prediction.
aIDT uses transfer functions based on the cross-interference of the incident illumination and first order scattering to recover the object's 3D structure~\cite{ling2018high}.
These transfer functions exhibit finite support and remove significant portions of the intensity measurements' nonlinear scattering signal that becomes non-negligible under multiple-scattering conditions~\cite{chowdhury2019high}.
This information loss limits the information available for the network to learn from, which contributes to the network's failure to predict object features including low axial spatial frequencies outside the linear model's bandwidth.
This limitation could be solved through the incorporation of higher order physical approximants~\cite{goy201821378}. 

Our learned IDT approach holds promise for improving the image quality in low-cost optical setups.
Recent works have developed low-cost, open-source optical imaging setups enabling affordable multi-modal imaging in a push for the ``democratization'' of science to the general public~\cite{cybulski2014foldscope, maia2017100, diederich2020versatile}.
Particularly, recent work from Diederich {\it et al.} has shown that aIDT can be included in such multi-modal setups enhancing both the capabilities of these platforms and accessibility to the imaging modality~\cite{diederich2020versatile}.
By using cheaper optical components, however, the reconstructed volume can suffer in quality from having less precision over the source array alignment.
Because our learned approach generalizes well to object recovery in different optical setups, the use of this lightweight framework on low-cost setups could drastically improve the object volume predictions and potentially be implemented on cellphones for real-time processing~\cite{diederich2019cellstorm}.
This generalizability could also correct the stronger model-induced artifacts present from multiplexed illumination schemes used for high-speed imaging~\cite{matlock_2019}.
Prior work showed that combining illuminations in each IDT intensity image trades image reconstruction quality and stronger missing cone artifacts for faster acquisition speeds to image live dynamic samples~\cite{matlock_2019}.
Since our network shows strong removal of these artifacts, this learned IDT approach could potentially be applied to multiplexed IDT setups to achieve faster volume rates without losing reconstruction quality.
This improvement would enable IDT to evaluate more dynamic biological features and enable high-speed imaging in low-cost optical setups as well.

The enhanced recovery of object RI features using our approach also has significant potential in recently developed digital staining technologies~\cite{christiansen2018silico,rivenson2019phasestain,kandel2020phase, cheng2021single}.
These methods showed that deep learning models coupled with the morphological features present in phase imaging modalities can synthesize artificial fluorescent labels on unlabeled cell samples.
This approach effectively adds specificity to the phase images for differentiating different biological structures and creates a ``computational multimodal'' setup expanding the phase modality's capabilities.
The enhanced 3D structures provided with this work could be utilized for such networks to provide computational specificity to the recovered RI volumes.
This avenue would provide a substantial boon for the biological imaging community by providing digital staining of 3D biological samples for analysis.

Finally, the generalizable network achieved using our physical model simulator shows the power of simulation-only training for applying  deep learning for imaging in complex media applications.
We showed here that the SSNP framework sufficiently models the multiple-scattering process for the network to recover experimentally measured samples from training only on natural image-based volumes. 
This learning approach overcomes the bottleneck of acquiring diverse experimental measurements for deep learning, and the model's efficient field simulation through large-scale multiple-scattering objects could be used for numerous other imaging advancements, such as imaging and light manipulation through diffuse media~\cite{li2018deep, li2018imaging, yoon2020deep, rotter2017light, rahmani2020actor, borhani2018learning}.

\section*{Materials and Methods}
\label{section:MM}
\subsection{aIDT experimental imaging setup}

The experimental aIDT setup consists of a commercial transmission microscope (E200, Nikon) equipped with an off-the-shelf LED ring (1586, Adafruit), a 0.65 NA, 40$\times$ objective (MRL00402, Nikon), and an sCMOS camera (Panda 4.2, PCO, pixel size 6.5$\mu$m).
The LED ring is placed $\approx 35$mm from the sample plane to generate oblique quasi-plane wave with illumination angles $NA_i ~\approx 0.63$.
Each aIDT measurement used for approximant generation included eight intensity images from a subset of the ring's 24 LEDs using green (515nm) illumination.
Each image was acquired with a 10ms exposure time providing a camera-limited 10.6 Hz acquisition speed for a single measurement.
Additional information regarding this setup and the post-processing calibration procedures are discussed in detail in~\cite{li2019high}.

\subsection{Training dataset generation}

Creating the training data for the learned IDT implementation required three steps: 1) generating sufficient large-volume object quantities with unique features, 2) simulating intensity images using a multiple-scattering forward model, and 3) recovering object approximants using the aIDT linear inversion model.
Unique object volumes were generated from randomly selecting $128\times128$ natural image patches obtained from multiple open-source databases including  SUN397~\cite{xiao2010sun}, Faces-LFW~\cite{huang2008labeled}, Caltech256~\cite{griffin2007caltech}, NWPU-RESISC45~\cite{cheng2017remote}, Stanford Online Products~\cite{oh2016deep}, CIFAR-10~\cite{krizhevsky2009learning}, and Stanford Dogs~\cite{khosla2011novel} datasets.
This random selection process was done from an equal selection of patches from each database to allow for equal probability of image type selection.
For each object volume, fifteen random selections were made with a 70$/$30 probability of selecting from the natural image list or a null slice containing no scatterers, respectively.
This process was empirically found to allow for sparse, weakly scattering samples and dense, multiple-scattering objects to be created simultaneously.
The final image volumes consisted of $128\times128\times15$ voxels with a voxel size of $0.1625\times0.1625\times0.9165 \mu m^3$.
The axial voxel size matches the microscope's depth-of-field (DOF) in water.
Following volume generation, the volume was normalized between $[-1, 1]$ and multiplied by a random RI value $n \in [0, 0.05]$ to generate the scattering object volume.
A background RI value matching water ($n=1.33$) was subsequently added to mimic evaluating biological samples in aqueous media.

The intensity images for each volume were generated using the recently developed SSNP multiple-scattering simulator~\cite{lim2019high} using version 3.6.9 of the Python programming platform.
For simulation, the generated volumes were padded with uniform values matching the background RI and the edges between this padding and the image were Gaussian filtered to reduce boundary artifacts.
The volume was remeshed axially to a size of $256\times256\times150$ voxels to provide smaller step sizes for generating valid SSNP simulations.
The simulation parameters used for SSNP matched the experimental aIDT setup, and a total of eight intensity images were generated for each object matching aIDT's illumination scheme.
The images were subsequently normalized and background-subtracted following the aIDT procedures~\cite{li2019high}.

The IDT linear inverse scattering model was implemented for recovering approximants of each object for training~\cite{li2019high}.
Fifteen transfer functions were generated to recover the original 15 simulated object slices with illumination angles, pupil size, imaging wavelength, and sample thickness matching the aIDT experimental setup~\cite{li2019high}.
Tikhonov regularization with a manually determined threshold value of 100 was implemented to recover each slice of the simulated object volume.
This regularization parameter, while traditionally chosen separately for each object based on an estimate of its signal-to-background (SBR) ratio~\cite{matlock_2019}, was fixed for this training set to provide examples of under-regularized and over-regularized objects to the network that can occur in practice due to user error.
This process of object generation, image simulation, and reconstruction was repeated until 10,000 training objects were obtained.
9,990 of these objects were used for training with five volumes reserved for validation and testing, respectively.
Additionally, a separate testing set of 1000 objects were generated using the same procedure with the previously unseen Food-101 dataset~\cite{bossard14} for evaluating the network performance in simulation without potential overfitting issues.


\subsection{Model-based linear fitting}

Prior to training the network. a model-based linear fitting approach was utilized to homogenize the input to the network.
The linear fitting process applies a Fourier transform to the object's intensity images and recovers the average magnitude of the eight intensity images' spectra within the maximum recovered bandwidth:
\begin{equation}
    \overline{|\hat{I}(\bm{\nu}) \odot P_{2NA}(\bm{\nu})|} = \alpha \Delta n_{max} + \beta, 
\end{equation}
where $\hat{I}(\bm{\nu})$ denotes the intensity spectra with spatial frequency coordinates $\bm{\nu}$, $\odot$ is the Hadamard product,  $P_{2NA}(\bm{\nu})$ is the circular bandwidth binary mask with radius 2NA (Supplemental), $\alpha$, $\beta$ are the linear coefficients, and $\Delta n_{max}$ is the true object's peak RI.
As shown in the supplemental material, this spectral magnitude behaves linearly with the peak RI value of the ground-truth object in the training dataset ($R^2$ = 0.917).
While the spectra is inherently nonlinear with RI, the result shows a dominantly linear behavior regardless of RI value enabling this normalization approach.
Using the linear fit, we normalize the input to the network using the estimated peak RI values based on the corresponding intensity spectral term.
This estimated RI scaling is subsequently applied to the network output. 
From this training data, the linear fit coefficients were found to be $\alpha=687.03$ and $\beta=-0.1591$ and used when normalizing both simulated and experimental data.

\subsection*{Network training and architecture}
Following normalization, the network was trained using using the simulated dataset for 100 epochs using the mean absolute error (MAE) loss function, ADAM optimization~\cite{kingma2014adam} with a $10^{-3}$ learning rate and batch size of 20.
The network used five consecutive randomly selected slices from the object approximants as inputs to predict the central slice of this subvolume.
This subvolume was randomly selected from the training data on each mini-batch during training to improve arbitrary volume generalization.
The loss and validation curves can be seen in the supplemental material.

\subsection*{Network architecture}
Our network follows a modified 2D U-Net architecture~\cite{ronneberger2015u} and was implemented using Tensorflow 2.0.0 with Keras.
It (Fig.~\ref{F2}) consists of five encoding and decoding layers with 128 initial filter channels and skip connections preserving high-resolution object features throughout the network. 
The encoding operation consists of two 2D convolution operations followed by a summation of their output feature maps in a residual connection.
The first 2D convolution provides downsampling using a $4\times4$ filter kernel size with a stride of 2, while the second convolution maintains feature size with a $3\times3$ kernel size with a stride of 1.
Both convolutions are followed by batch normalization and a Leaky Rectified Linear Unit (LReLU) activation function to allow prediction of negative RI contrast~\cite{xu2015empirical}.
At the bottleneck, the network uses an additional 2D convolution with a $1\times1$ filter kernel with a stride of 1 for blending the latent feature information learned from the 3D input.
The resulting feature map is decoded using a process of $2\times2$ upsampling followed by a $4\times4$ 2D convolution with a stride of 1 and two sets of 2D convolution operations with $3\times3$ convolution kernels also with a stride of 1.
Each convolution operation is followed by batch normalization and LReLU.
This design recovers the original image size without incurring checkerboard artifacts from transpose convolution operations and provides sufficient filters for learning image features.
For the final prediction, the network performs a 2D convolution with $1\times1$ kernels without a nonlinear activation to predict the central object slice.

\section*{Funding} 
National Science Foundation (1846784). Alex Matlock acknowledges the National Science Foundation Graduate Research Fellowship (DGE-1840990).

\section*{Acknowledgment} 
The authors would like to thank Jiabei Zhu for implementing the SSNP algorithm.

\section*{Disclosures} 
The authors declare no conflicts of interest.

\section*{Supplemental document} 
See Supplement 1 for supporting content.

\bibliographystyle{IEEEtran}

\end{document}